# Roughening transition as a driving factor in the formation of self-ordered one-dimensional nanostructures

Vyacheslav N. Gorshkov[1], Vladimir V. Tereshchuk[1], and Pooya Sareh[2*]

[1] National Technical University of Ukraine, Igor Sikorsky Kyiv Polytechnic Institute,
37 Prospect Peremogy, Kiev 03056, Ukraine.

[2] Creative Design Engineering Lab (Cdel), School of Engineering, University of Liverpool, Liverpool, The Quadrangle, Brownlow Hill, L69 3GH, United Kingdom. *Corresponding author. Email: pooya.sareh@liverpool.ac.uk

## Abstract

Based on the Monte Carlo kinetic method, we investigated the formation mechanisms of periodical modulations arising along the length of one-dimensional structures. The evolution of initially cylindrical nanowires/slabs at temperatures lower than their respective melting temperatures, can result either in breakup into single nanoclusters or in the formation of stable states with pronounced modulations of cross section. Such configurations, observed in a number of experiments, are excited at wavelengths $\lambda$, which are below the critical value, $\lambda_{cr} = 2\pi r_{nw}$, for the development of classical Rayleigh instability ($r_{nw}$ is the initial radius of the nanowire). We show that the modulation excited in the "subcritical mode", $\lambda < \lambda_{cr}$, corresponds to the appearance of roughening transition on the quasi-one-dimensional surface of nanowires/slabs, which is responsible for the transformation of smooth metal facets (two-dimensional systems) at a temperature, $T$, which exceeds a certain critical value, $T > T_R$.

Since the arise of roughening transition is possible only on certain facets of metals with a given crystal structure, the short-wavelength modulations of one-dimensional systems, as shown in our work, can be realized (i) with the proper orientation of the nanowire/slab axis providing spontaneous appearance of roughening transition on its lateral surface, (ii) by the method of activating the surface diffusion of atoms by external impact (irradiation with an electron beam or contact with a cold plasma), which stimulates roughening transition without significant heating of the nanowire.

For the cases of BCC and FCC lattices, we have demonstrated that it is possible to excite either metastable structures with a wavelength below the threshold of energetic instability, $\lambda < 4.5 r_{nw}$, or substantially increase $\lambda$ ($\lambda \gtrsim 12 r_{nw}$,) by controlling the distance between nanodroplets during the nanowire breakup. The results obtained can be used in the controlled synthesis of ordered one-dimensional structures for use in optoelectronics and in ultra-large-scale integrated circuits.

# 1. Introduction

Numerous beneficial properties of nanowires have inspired and promoted their extensive study in recent years. The low resistivity of gold nanowires makes them almost perfect 1D conductors, enabling these structures to be used as interconnects [1], while, due to high stability, tungsten nanowires are indispensable in such applications as smart coatings [2] and lithium-ion batteries catalysts [3]. At the same time, low reflectance [4,5], large intrinsic carrier mobility [6-9], and high light-harvesting capacity [10] of semiconductor nanowires are advantageous for sensing, optoelectronic and photovoltaic applications [11-17]. It should be also noted that existing methods of synthesis allow tuning the surface morphologies of nanowires to the unprecedented range [18-26], which makes them also promising for optomechanical studies [28,29].

However, due to poor thermal stability, the implementation of the aforementioned beneficial characteristics remains challenging. The morphological instability of nanowires under elevated temperatures significantly impairs their optoelectrical properties. For instance, at large current densities, the Joule heating is known to cause the formation of surface perturbations on the nanowire and subsequently lead to interconnect failure. It should be noted that the thermal instability of nanowires has been the subject of discussions over the last two decades [30-55].

The development of periodic perturbations of the surface of a cylindrical nanowire (with a wavelength of $\lambda$) is accompanied, according to an approximate analytical model [31], by a decrease in its surface energy, $E_s$. If the surface energy density, $\sigma$, is isotropic (i.e. it does not depend on the orientation of the surface element relative to the internal crystalline structure), then a decrease in $E_s$ is associated only with a decrease in the lateral surface area, $A_{nw}$, which is possible only at $\lambda > \lambda_{cr} = 2\pi r_0$ ($r_0$ is the initial radius of the nanowire). The change in the configuration of the wire is due to the surface diffusion of atoms from the neck regions of the cross section to the zones of broadening.

The optimal ratio of the mass of transported matter to the change in surface energy (maximum instability increment) is achieved at $\lambda_{max} \approx 9r_0$. In the described model [31], the dynamics of the nanowire is very similar to the results of the classical theory of instability of Plateau-Rayleigh liquid jets [30]. A recent work [48] took into account the possible anisotropy of $\sigma$ under the assumption that the shape of a nanowire can also be presented by sinusoidal modulation of its radius along the length. The obtained qualitative results are physically transparent. If the value of $\sigma$ in the narrowing region is less than the same value on the lateral surface, then instability develops at shorter wavelengths, $\lambda < 9r_0$. For an inverse ratio, a decrease in $E_s$ can be achieved only by significantly reducing $A_{nw}$, which corresponds to wavelengths $\lambda > 9r_0$. The experimentally observed deviations of $\lambda$ [20-22, 25, 32, 38, 39, 53] on both sides of $\lambda_{max}$ are largely associated with a different type of anisotropy $\sigma$, which depends on the orientation of the nanowire relative to its crystal structure [45-48,53].

However, a number of experiments are known in which significant modulations of the nanowire radius with wavelengths $\lambda < 2\pi r_0$ **[21-22,25,32,53]** are observed. On the one hand, the possible breakup of a nanowire into short-wavelength fragments does not contradict the so-called "energetic instability", when the surface energy after decay is less than its initial value. For an isotropic value of $\sigma$, this relation is realized for $\lambda > 4.5 r_0$ (the area of the lateral surface of an infinite nanowire exceeds the total surface area of spherical nanodroplets formed from it). On the other hand, the transition from a state with a higher potential/surface energy to a lower one requires overcoming some intermediate energy barrier, which cannot be realized according to the dynamics equations describing the development of nanowire instabilities in the linear approximation **[52]**.

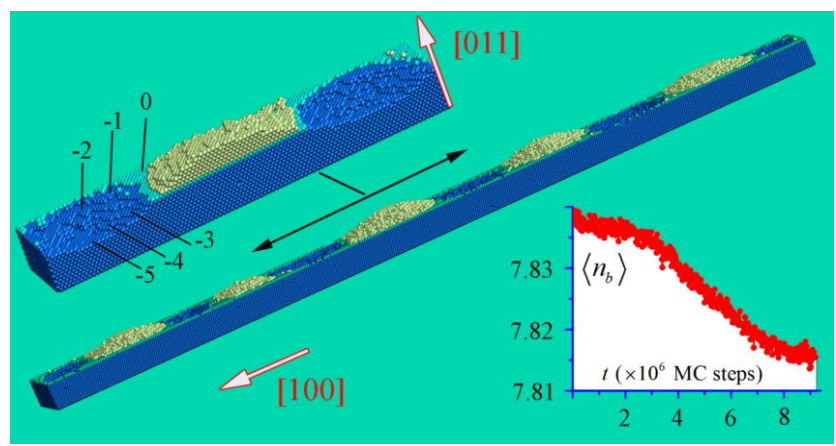

**Fig. 1 The manifestation of roughening transition on the (110)-facet of the plate that is a small near-surface part of a bulk metal**. Only the atoms that form this plate are supposed to be movable while the ambient atoms are "frozen"/moveless (the movable atoms cannot come out of the perimeter of the upper facet). The dimensions of a slab are as follows: length $L = 630$, width $w = 70$, and thickness $h = 20$. Hereinafter, the unit of length is the distance between the neighbor (100) atomic layers, $a/2$ ($a$ is the lattice constant). The number of slab atoms is equal to $N_0 \approx 232 \times 10^3$. The upper insert shows an enlarged image of the part of the slab. Layers colored in yellow and blue depict the formed stepped regions above and below the initial surface/"zero" layer (colored in green) of the slab, respectively. The insert in the lower right corner demonstrates the dependency of the average number of bonds per atom, $\langle n_b \rangle$, on time.

Nevertheless, as it is mentioned above, the excitation of short-wavelength structures is the really observed phenomenon. Moreover, the developed pronounced modulations of the nanowire radius sometimes become "frozen" in time. In these cases, analogues of the equilibrium configurations of radius-modulated liquid jets (unduloids) are realized, which exist for $2\pi r_0 > \lambda > 4.5 r_0$ **[21,22,32]**. That is, not every process of breakup of a nanowire into fragments can be interpreted on the basis of analogies with Rayleigh instability **[30]**, otherwise one can come to the erroneous conclusion of possibility of either stabilizing the breakup or controlling its main parameters.

Our work shows the manifestation of one of the factors that can significantly modify the mechanism of spontaneous periodic modulation of a nanowire cross section up to its breaking into fragments. Such a factor is the roughening transition effect **[56-66]**. If the temperature of a flat surface, $T$, is higher than critical, $T > T_R$, then this surface becomes periodically-modulated in height with an increase in its area. Such a transformation at a fixed temperature (even with an increase in surface energy) does not contradict the laws of thermodynamics, since in this case the free energy, $F$, of the system decreases: $dF = dU - TdS < 0$, where $U$ is the internal energy and $S$ is the entropy of the system. Roughening transition has been studied in detail for silicon **[47,63,64]**. It manifests itself at different temperatures depending on the orientation of the substrate plane and is associated with an intensive exchange of its surface with the near-surface layer of atoms evaporated from it.

If the roughening transition occurs on flat surfaces, then it should also develop on the lateral surface of nanowires, as demonstrated in our numerical experiments. Cases of nanowires with FCC and BCC crystal lattices are considered. The first of them was initiated by the results with gold nanowires **[32]** with a radius of the order of $\approx 7 nm$. Under the influence of electron beam irradiation, modulations of the radius with the wavelength $\lambda/r_0 \sim 5.2 - 5.6$ appeared in accordance with our estimates of the data given in **[48]**. We attribute this to the stimulation of the roughening transition in electron bombardment of a nanowire surface. In addition, the perturbations of a radius of noticeable amplitude, which arose in the first 10 minutes, stopped growing despite further irradiation. The Monte Carlo model we use also detects this phenomenon. The study of the decay of nanowires with a BCC lattice was stimulated by **[20]**, which considered the problem of stabilization of tungsten filaments.

At a given temperature, the roughening transition occurs on selected faces of the crystalline structure of a material. In Fig. 1, the results of our numerical model demonstrate the manifestation of this effect on the (110)-type surface of a plate. (This plate ($L \times w \times h$) presents the near-surface layers of the bulk metal with body-centered cubic lattice only within the surface region ($L \times w$)). A decrease in time of the average number of bonds per atom, $\langle n_b \rangle$, with its nearest neighbors reflects an increase in its internal energy, $U$. Nevertheless, the observed increase in the total surface area corresponds to such an increase in entropy, $S$, that $dF = dU - TdS < 0$. The only dimensional parameter with which it is logical to compare the length of the developed perturbations is the slab width, $w$ (the number of formed hillocks equals to 5-6 in our simulations). However, the value $\frac{\lambda}{w} \approx 1.5 - 1.8$, in this case, is not directly related to the physical factors determining the observed periodic disturbances, and the process of their excitation cannot be considered as the Rayleigh instability.

The initial stage of nanowire annealing is accompanied by the transformation of its initially cylindrical surface into a faceted surface composed of a set of various crystalline faces. The faces that undergo the roughening transition stimulate the development of perturbations with a wavelength that are inherent in this process. On other faces, the surface perturbations can develop according to their physical laws. Therefore, the scenario of the entire breakup process can be very ambiguous as a result of competition between different physical mechanisms. When the roughening transition dominates, the $\lambda/r_0$ ratio can be observed noticeably below the known limit ($2\pi$). The development of such relatively short-wavelength perturbations, as shown in our work, can lead to the formation of periodically modulated one-dimensional structures "frozen" in time at the nonlinear stage of interaction of disturbance modes.

Note that metal nanowires with a BCC structure have a special feature in the formation of a side surface at the initial stage of annealing. In the [100] and [111] orientations, their lateral surface is mainly represented by (110) faces with the minimum surface energy density, which are most susceptible to the roughening transition (see Fig. 1). The mentioned property determines the strong dependence of the breakup characteristics into nanodroplets depending on the orientation of the axis of the nanowire relative to the crystal structure. In addition, the degree of manifestation of the roughening transition in one-dimensional nano-systems (not only in nanowires, but also in nanoribbons) can change during phase transitions associated with a change in the symmetry of its crystal structure due to external influences **[49]**. In particular, the effect of periodic modulation of nanoribbons (Au) is stimulated by their irradiation with an electron beam **[49]**, and highly ordered Ag particle-chains have been obtained from Ag nanowires by radio frequency Ar$^+$-plasma treatment **[50]**.

In our work, we concentrate on the study of factors determining the transformation of one-dimensional extended nanosystems, which go beyond the established conceptions of the mechanisms of their instability **[31]**. The numerical experiments that we performed on the basis of the Monte Carlo mesoscopic model are focused on studying the most general mechanisms of the formation of periodically modulated nanostructures without focusing on specific metals.

## 2. Kinetic Monte Carlo model

The applied model (developed by one of the co-authors of this work) was successfully used in previous studies on the diffusion growth of nanoparticles with different shapes from the same material **[67-69]**, sintering of nanoparticles embedded in a polymer paste **[70,71]**, and synthesis of ordered systems of nano-pillars on the surface of a substrate for catalysis applications **[72,73]**. Based on this model, the processes of breakup of nanowires with the FCC and diamond-like lattice structures were investigated **[45-48]**. Detailed description of the model is presented in the works mentioned above. Here we present only its basic conceptions.

The model assumes that the atoms of the nanowire/nanoribbons are located in the nodes of the crystal lattice of a given type. It uses two parameters: the first one, $\alpha$, reflects the energy of pair interaction, $\varepsilon < 0$, of neighboring atoms:

$$\alpha = |\varepsilon|/kT, \qquad (1)$$

and the second parameter, $p < 1$, determines the mobility of atoms and depends on the energy/activation barrier, $\Delta$:

$$p = \exp(-\Delta/kT), \tag{2}$$

where $T$ is temperature.

The dynamics of the nanosystem is presented in the sequence of Monte-Carlo (MC) steps. Each of these steps involves the following operations. If there are $N_0$ atoms in the system, then the same number of times we randomly select one of them (on average, once per MC step) and determine its possible new position. If this atom has $n_{vac}$ nearest vacancies (unoccupied lattice sites), then the probability of an attempt to jump, $p_{jump}$, into one of them is

$$p_{jump} = p^{m_0}, \tag{3}$$

where $m_0$ is the number of nearest occupied lattice sites.

When implementing the jump, the new state of the atom is selected from $(n_{vac} + 1)$ candidates (including the initial state), with the probability, $p_{target}$, of each of them being proportional to

$$p_{target} \sim \exp(m_t|\varepsilon|/kT), \tag{4}$$

where $m_t$ is the number of nearest neighbors in the supposed new state. That is, the end position is selected according to the set of Boltzmann factors.

The presented algorithm allows the sublimation of some bound atoms from the surface of the nanocluster. Thus, if the selected atom turns out to be free, a random-direction diffusion hop is carried out with fixed-length step, $\ell$, that is a fraction of the lattice constant, $a$. This atom becomes crystal-lattice registered as part of the nanostructure if its final position is within a unit cell near a crystal. The sublimation process can be blocked and then only the surface diffusion of bonded atoms determines the nanocluster dynamics.

Here we also note some technical details. The ends of the nanowire are in contact with several atomic layers (~ 5) composed of motionless ("frozen") lattice atoms. This trick is up to some extent equivalent to periodic boundary conditions. An extended one-dimensional nanocluster is enclosed in a cylindrical container, the walls of which reflect free atoms falling on it. The radius of this container is approximately an order of magnitude greater than the characteristic transverse size of the nanocluster under study.

The results of previous numerical studies **[67-73]**, which are in good agreement with experimental data, were obtained using the so-called reference values $\alpha_0$ and $p_0$ comparable to 1. Changes in temperature, $T$, entail changes in the parameters $\alpha$ ($\alpha \sim 1/T$) and $p$, which are related to each other by the equation

$$p = (p_0)^{\alpha/\alpha_0}. \tag{5}$$

According to the aforementioned studies, we use the following reference "intermediate temperature" values: $\alpha_0 = 1.5$ and $p_0 = 0.65$ for the BCC crystal lattice; and $\alpha_0 = 1.0$ and $p_0 = 0.7$ for the FCC lattice.

Note that when modeling the dynamics of one-dimensional nanoclusters with a BCC structure, we take into account the interaction of atoms only with their nearest neighbors, although in some cases, as stated in **[74-77]**, interactions with first- and even second-order neighbors can make some contribution. We do not complicate our numerical model, taking into account these possible interactions, since, as mentioned above, the goal of this study is to show the unusual scenarios that can arise in the dynamics of one-dimensional nanosystems, which are difficult to interpret based on the traditional conceptions of Rayleigh instability. This especially concerns to nanowires with a BCC crystal lattice due to the specific morphology of the Wulff construction in this case (see Fig. 7).

## 3. Results

### 3.1 Dynamics of nanowires with BCC lattice structure

As we noted above, the features of the breakup of nanowires are associated with the degree of anisotropy of the surface energy density $\sigma$. In our model, we take into account the interaction of lattice atoms only with their nearest neighbors. The relationship between surface energy densities on faces with small Miller indices (i.e. (100), (110), and (111)) can be easily established by calculating the ratio $n_{br}/A_s$ ($\sigma \sim n_{br}/A_s$), where $n_{br}$ and $A_s$ are the numbers of broken bonds and part of the face area per one surface atom, respectively **[74-76]**.

For the BCC lattice under consideration, $n_{br}^{(100)} = 4$, $n_{br}^{(110)} = 2$, $n_{br}^{(111)} = 6$, with $A_s^{(100)} = a^2$, $A_s^{(110)} = a^2/\sqrt{2}$, $A_s^{(111)} = \sqrt{3}a^2$ ($a$ is the lattice constant). Thus, the minimum value of $\sigma$ is achieved on the facet (110) and

$$\sigma_{(110)} : \sigma_{(111)} : \sigma_{(100)} = 1 : \sqrt{1.5} : \sqrt{2} \approx 1 : 1.22 : 1.41 \qquad (6)$$

The possible contribution of the nearest six second-order neighbors, which is characterized by the value of the parameter $\rho$ **[74]**, gives the following relations

$$\sigma_{(110)} : \sigma_{(111)} : \sigma_{(100)} \approx 1 : 1.22 : 1.34 \text{ (for W and Mo, } \rho \approx 0.11\text{)}$$

$$\sigma_{(110)} : \sigma_{(111)} : \sigma_{(100)} \approx 1 : 1.22 : 1.30 \text{ (for V, } \rho \approx 0.18\text{)} \qquad (7)$$

$$\sigma_{(110)} : \sigma_{(111)} : \sigma_{(100)} \approx 1 : 1.23 : 1.30 \text{ (for } \alpha-\text{Fe}, \rho \approx 0.2\text{)}$$

The value $\rho$ is the ratio of the second neighbor bond energy to the nearest bond energy. One can see that, with the approximation of taking into account only the nearest neighbors, the model error may be insignificant.

The facets with minimum surface energy are the least distant from the center of the nanoparticle that takes the form of Wulff construction in its equilibrium state. In the case of BCC lattice, this equilibrium configuration is rhombic dodecahedron (see Fig. 7) all 12 faces of which are represented by a set of ⟨110⟩-planes. It is easy to see that a quasi-one-dimensional nanostructure that is elongated along the [100]-axis can be bounded by four faces of the (110) type and will have a minimum surface energy. It is this form that the nanowire takes in the initial stage of its transformation (see inset B of Fig. 2; recall that theunit of length is half the lattice constant $a$), and it is the (110)-facets at which the roughening transition develops, as demonstrated in Fig. 1.

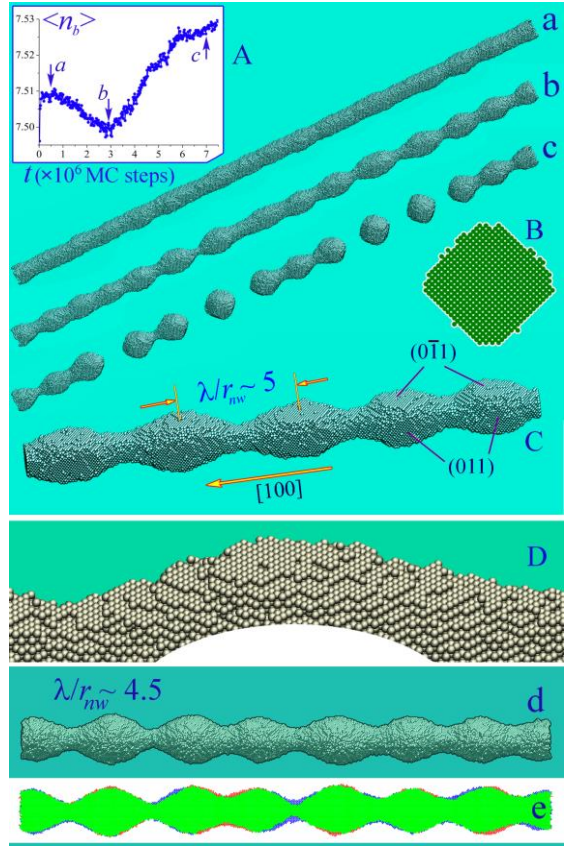

**Fig. 2**. **Dynamics of a [100]-nanowire breakup at different temperature regimes**. Sub-images (a), (b), and (c) – taken at moderate temperature: $\alpha = 1.5$ and $p = 0.65$; $L = 1000$, diameter $d_0 = 30$ (~4.7 nm for W), and $N_0 \approx 174.7 \times 10^3$ – depict the shapes of nanowire at time moments $t = 1, 3,$ and $7 (\times 10^6)$ MC time steps, respectively. Inset A shows the dependency of the average number of bonds per atom on time. Inset B presents the cross-section of the nanowire at the initial stages of breakup. Inset C shows a fragment of the nanowire given in configuration (b). Insert D represents a part of the stepped bulge surface as a set of terraces formed as a result of thermal roughening. Sub-images (d) and (e) − taken at warm regime: $\alpha = 1.3$ and $p = 0.69$; $L = 630$, $d_0 = 40$, and $N_0 \approx 199.7 \times 10^3$ − represent the shapes of nanowire at time moments $t_1 = 4.3 \times 10^6$ and $t_2 = 13.1 \times 10^6$ MC time steps. Configuration (e) is the overlay of the nanowire projections onto the (010)-plane. Green and blue regions show the shape of nanowire at $t_1$, whereas green and red regions represent its shape at $t_2$. Evaporation is blocked in both cases.

An increase in the amplitude of the modulations of the nanowire surface in time (see configurations $a$ and $b$ in Fig. 2) is accompanied by a decrease in the average number of bonds $\langle n_b(t) \rangle$ (see inset A in Fig. 2), which is analogous to the results shown in Fig. 1. A decrease in the surface energy of the wire (an increase in the parameter $\langle n_b(t) \rangle$) occurs only during breakup into individual nanoclusters, when the area of its lateral surface decreases sharply. Inset D in Fig. 2 demonstrates the formation of terraces from the fragments of (110)-facets. The pronounced similarity in the physical mechanism of evolution of the plate and nanowire surfaces (see Fig. 1 and Fig. 2) is completed by a small (below the critical value $2\pi r_{nw}$) value of the decay parameter $\lambda/r_{nw} \sim 5$ (Inset C in Fig. 2). However, we note that the value of this ratio changes with variations in the radius of nanowire due to the physical features of surface transformation, which we will consider below.

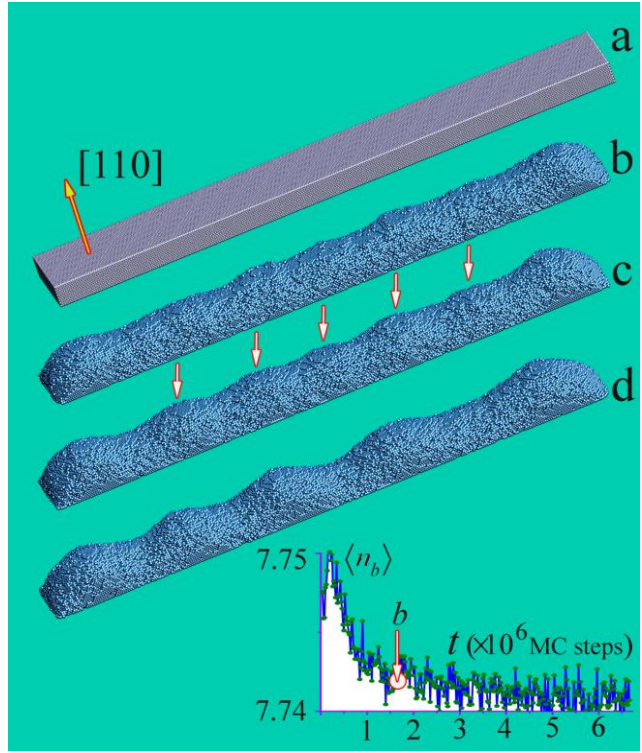

**Fig. 3. Dynamics of a slab lying on a substrate of moveless atoms.** $\alpha = 1.3$ and $p = 0.69$; $L = 560$, $h = 20$, and $w = 80$; $N_0 \approx 240 \times 10^3$. (a), (b), (c), and (d): $t = 0, 1.6, 2.6, 6.7$ ($\times 10^6$ MC steps). The inset shows the dependence of the average number of bonds, $\langle n_b \rangle$, on time; the circle marks the moment when the slab takes the configuration (b).

Terraces formed above the upper layer of the plate (0-layer), shown in Fig. 1, is caused by jumps of atoms from this 0-layer to the initially unfilled +1-layer. If the intensity of such transitions is sufficiently high, the atoms of the +1 layer, drifting along the 0-layer, manage to form unified clusters, which can be the basis for the formation of similar clusters/terraces in the +2, +3-layers, etc. As a result, in Fig. 1, we can see many distinct stepped structures equidistant from each other. Recall that in our model we take into account the interaction of only the nearest neighbors, therefore, the observed long-scale ordering of these structures is unexpected. The mechanism that determines this ordering is the surface diffusion of atoms, which provides a kind of long-range interaction between the forming stepped structures.

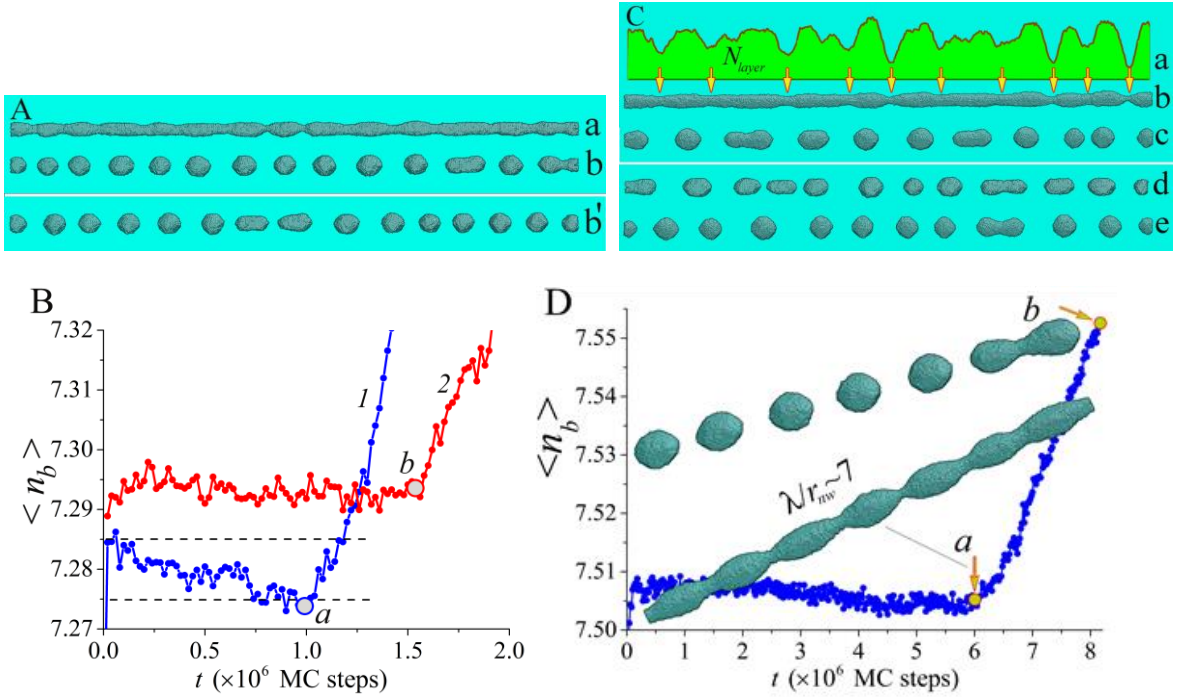

**Fig. 4**. **Disintegration of a [100]-nanowire at moderate temperature**: $\alpha = 1.5$ and $p = 0.65$. (A) $L = 1000$ and $d_0 = 20$. The evaporation is blocked; $\langle\lambda\rangle/r_{nw} \approx 7.1$. Sub-images (a) and (b) depict the system at $t = 1 \times 10^6$ and $2 \times 10^6$ MC steps, respectively; sub-image (b') shows the result for another random MC simulation at $t = 2.4 \times 10^6$ MC steps. (B) Characteristic dependencies of the average number of bonds per atom, $\langle n_b(t)\rangle$, when evaporation is blocked (curve 1) and is turned on (curve 2). The corresponding nanowire configuration for point $a$ is shown in frame (A)-(a), and that for point $b$ in frame (C)-(b). (C) Break-up of a nanowire when evaporation is taken into account. System parameters are the same as in (A); $\lambda/r_{nw} \approx 9.5$. Sub-images (a) and (b) show the dependency of number of atoms in (100)-atomic layers along the nanowire, $\widehat{N}_{layer} = N_{layer}/\langle N_{layer}\rangle$, and its configuration at $t = 1.56 \times 10^6$ MC steps; sub-image (c) depicts the final stage of the nanowire breakup, $t = 2.9 \times 10^6$ MC steps; sub-images (d) and (e) present the results of two another random MC simulations, $t = 2.4 \times 10^6$ and $t = 3.58 \times 10^6$ MC steps, respectively. (D) Dependency $\langle n_b(t)\rangle$ for a long nanowire with radius $r_{nw} = 15$. The evaporation is turned on. In the insets, the fragments of this nanowire corresponding to points $a$ and $b$ are shown.

It is known that the roughening transition effect develops at a temperature exceeding a certain threshold value $T_R$ ($T > T_R$) and provides the necessary intensity of the transition of atoms from at least the 0-layer to the +1-layer. Results shown in Fig. 1, and parts (d) and (e) of Fig. 2, were obtained in the temperature regime, which we call "warm": $\alpha = 1.3$ and $p = 0.69$ (see Eq. (5)). A numerical experiment with the same plate but with moderate heating ($\alpha = 1.5$ and $p = 0.65$) did not lead to the formation of ordered stepped structures of noticeable height (i.e., the corresponding temperature $T < T_R$). However, at the same temperature, a nanowire that is mainly bounded by the (110)-facets, may break up into fragments (see configurations (a), (b), and (c) of Fig. 2) in a relatively short time in comparison with the evolution time of the slab presented in Fig.1. We attribute this effect to the diffusion flux of atoms from the edges, which are formed by adjacent (110)-facets of the lateral surface of the wire (see inset B in Fig. 2). This flow feeds the weak supply of atoms from the side (110)-facets of the wire to the adjacent +1-layers. To confirm what has been stated, we present an experiment with a slab (Fig. 3), which lies on a substrate of "frozen" atoms (the side faces of the plate are free, unlike the variant shown in Fig. 1). From the above data, the two-stage formation of periodic structures on the surface of the plate due to the roughening transition is observed (Fig. 3). During the first stage, the diffusion outflow of atoms from its upper side ribs over the 0-layer leads to the appearance of short-wave modulations with a wavelength of $\lambda_1 \approx 560/7 = 80$ (see configurations (b) and (c) in Fig.

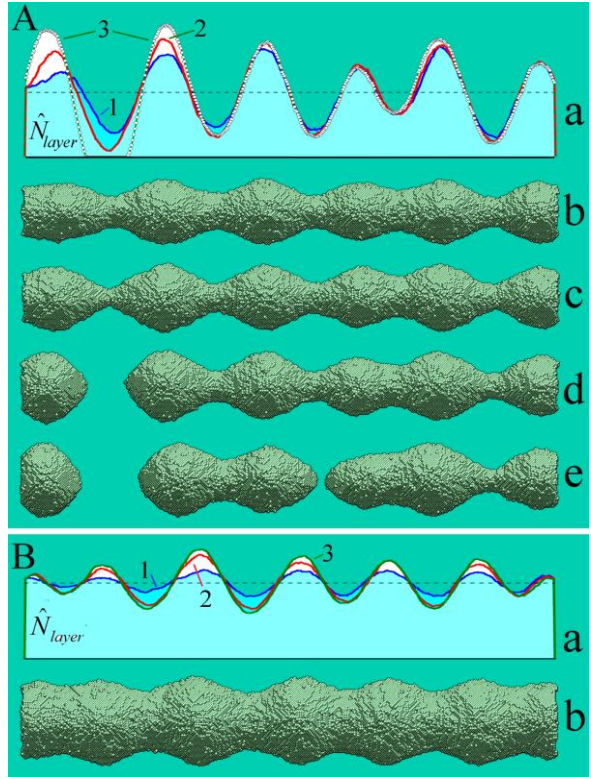

**Fig. 5.** **Dynamics of thick nanowires at moderate temperature:** $\alpha = 1.5$ and $p = 0.65$; $L = 470$. (**A**) $d_0 = 40$. (a) – Distributions of $\widehat{N}_{layer}$ along the nanowire at different times: $t = 6, 12, 15 \,(\times 10^6)$ MC steps for 1(blue), 2(red) and 3(olive) curves, respectively ($\langle N_{layer} \rangle = 316$). Configurations (b), (c), and (d) present the nanowire shapes at the abovementioned times and at $t = 30 \times 10^6$- configuration (e). (**B**) $d_0 = 60$. (a) – 1(blue), 2(red) and 3(olive) curves present the distributions of $\widehat{N}_{layer}$ along the nanowire at times: $t = 3, 5, 7 \,(\times 10^6)$ MC steps, respectively ($\langle N_{layer} \rangle = 712$). The configuration (b) shows the nanowire shape at time $t = 7 \times 10^6$ MC steps.

3), accompanied by a sharp decrease in $\langle n_b(t) \rangle$. Then follows the process of slow relaxation of the short-term initial perturbations when its long-wave mode is formed. As a result, - a chain of stepped structures visible in configuration (c) in Fig. 3 are absorbed by adjacent hillocks (see configuration (d)). At this stage, $\lambda_2 \approx 560/5 = 112$, which is close to the result shown in Fig. 1. Obviously, a correlation of the wavelengths $\lambda_1$ and $\lambda_2$ with the characteristic size, $w$, of the plate does not have physical meaning.

Note that the perturbation wavelength, $\lambda \approx 90$, in configuration (d) (see Fig. 2) approaches the value of $\lambda$ obtained at the initial stage of transformation of the slab presented in Fig. 3 ($\lambda_1 \approx 80$). Since the ratio $\lambda/r_{nw} \approx 4.5$ is close to the energetic instability threshold, the shape of the nanowire becomes practically frozen in time and represents an analog of the periodic equilibrium structure (unduloid) known for a cylindrical liquid jet **[78]**. In a liquid, such a state is unstable and ends either with a jet rupture or a transition to the initial state with weak noise modulation of the radius. There is only one evolutionary path for a wire - its breaking. In this case, the time to rupture, $t_{br}$, can be quite long. In the considered case, $t_{br} > 13 \times 10^6$ MC steps, which is noticeably longer than the formation time of the unduloid-like structure (see configuration (e) in Fig. 2).

A significant role of roughening transition in the decay of a nanowire introduces certain inertia of changes in the perturbation period, λ, with variations in its radius. The breakup of a nanowire with radius $r_{nw} = 15$ (Fig. 2, configurations (a), (b) and (c)) corresponds to a wavelength of $\lambda \approx 75$ ($\lambda/r_{nw} \approx 5$). For a wire of a smaller radius (see Fig. 4A, $r_{nw} = 10$) $\lambda \approx 71$ ($\lambda/r_{nw} \approx 7.1$) with the same parameters $\alpha$ and $p$. That is, the wavelength of the perturbations is almost the same, because it is caused by the same mechanism of their formation, which is very indirectly associated with the characteristic transverse size of the nanowire.

At the same time, if the indicated instability parameters were obtained in a real experiment ($\lambda/r_{nw} \approx 5$ and $\lambda/r_{nw} \approx 7$), then the generality of the breakup mechanism in these cases would seem clearly doubtful. Note that this generality is indicated not only by the identical wavelengths of surface disturbances, but also by the similarity in the dependences $\langle n_b(t) \rangle$ – in both cases, this parameter decreases by approximately the same value until the beginning of the first ruptures (see inset A in Fig. 2 and curve 1 in Fig. 4B).

The effects of roughening transition are most pronounced on thicker [100]-oriented nanowires. The results presented in Fig. 5 are obtained for a moderate temperature regime in which, as we noted above, the modulations of the (110)-plane do not develop. However, nanowires representing quasi-one-dimensional systems, which are bounded mainly by adjacent (110)-faces, are still subject to periodic disturbances in their cross section. We note a weak dependence of the wavelengths of these perturbations on the radius of the nanowire. In both variants presented in Fig. 5: $\lambda \approx 80$. The nearness of this value to the estimation of the perturbation wavelengths $\lambda_1$ observed at the first stage of plate evolution (see Fig. 3) was noted by us earlier in the analysis of results of Fig.2. That is, in all the cases under consideration, the outflow of atoms from interjacent zones connecting adjacent (110) faces is the determining factor, which stimulates the formation of stepped configurations on these facets. (The interjacent zones are formed by narrow ribbons from the planes of (100)-type, which, as shown below, are also subject to the occurrence of roughening transition).

Since at large radii of the nanowire the ratio $\lambda/r_{nw}$ falls below a critical value, $\lambda < \lambda_{cr} = 4.5 r_{nw}$, the resulting periodic surface modulations are in a metastable state ($\lambda/r_{nw} \approx 4.2$ in Fig. 5A and $\lambda/r_{nw} \approx 2.7$ in Fig. 5B). The distributions of $\widehat{N}_{layer}$ along nanowires (see sub-images (a) in Figs. 5A and 5B) demonstrate the formation of these states. Such "frozen in time" configurations are observed in many experiments with nanowires from different materials. In addition to the "natural origin" of metastable structures, their appearance can be stimulated by the external conditions which impact the nanowire surface (see the next section). The breakup time of a thick nanowire into individual fragments is highly random and inevitably accompanied by the merger of neighboring beads into larger nanoclusters. Fig. 5A shows the result of a numerical experiment when a nanowire rupture occurred in a relatively short time. However, even after the first breakup, the dynamics of its longer (right) fragment are almost invisible.

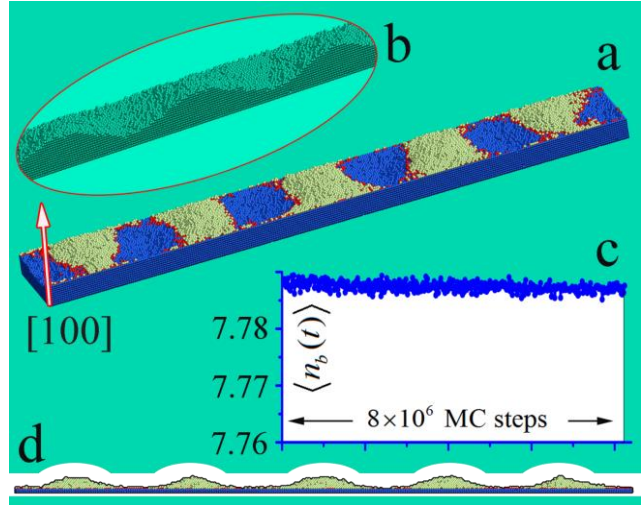

**Fig. 6. Manifestation of roughening transition on the upper (100)-face of a slab.** Warm regime: $\alpha = 1.3$ and $p = 0.69$; $L = 550$, $h = 20$, $w = 70$, and $N_0 \approx 205 \times 10^3$. The lateral surface of the slab (with the exception of the upper face) is in contact with "frozen" atoms. Evaporation is blocked. Sub-images (a) and (d) show the shape of the slab in detail at $t = 8 \,(\times 10^6)$ MC time steps. Inset (b) presents the formed hillocks and cavities on the slab. Inset (c) represents the dependence of the average number of bonds per atom, $\langle n_b \rangle$, on time.

In some cases, the evaporation of atoms from the surface of a nanowire is possible. Then its instability develops under the conditions of the exchange by atoms between the nanowire surface and the near-surface layer of free atoms. A detailed analysis of the effect of such an exchange was carried out in **[47]** when studying the mechanisms of the breakup of nanowires with like-diamond crystal structure. The results obtained in **[47]** show that, when evaporation is taken into account, the length of surface perturbations can decrease and fall below the known classical limit – $\lambda < \lambda_{cr} = 2\pi r_{nw}$. In our case of the BCC lattice, the opposite effect is realized (see parts C and D of Fig. 4). The transfer of atoms within the surface layer weakens the role of surface diffusion in the development of instability – the decrease in $\langle n_b(t) \rangle$ is very weak (see curve 2 in Fig. 4B), and the value of $\lambda$ exceeds the critical value $\lambda_{cr}$.

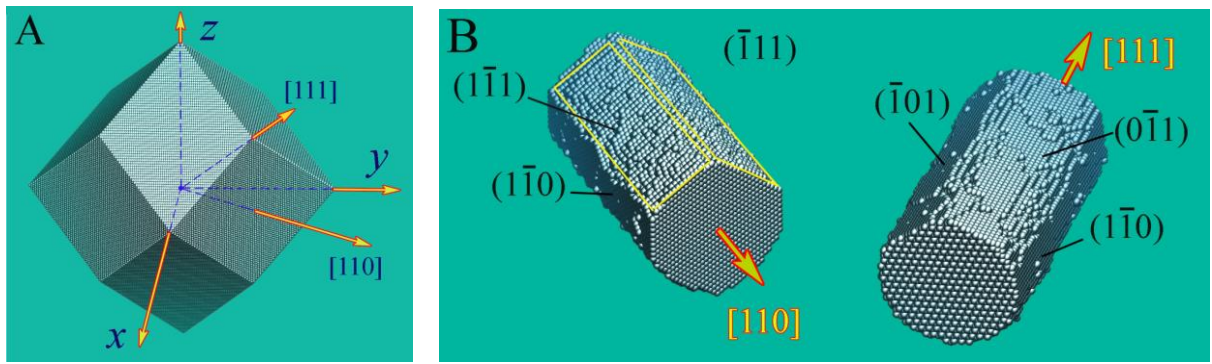

**Fig. 7.** **(A)** The Wulff construction for the BCC crystal lattice. **(B)** The structure of the lateral surface of nanowires with [110]- and [111]- orientations at the initial stage of evolution ($\alpha = 1.5$ and $p = 0.65$). The parameters of the nanowire fragments are as follows: $L = 120$, and $d_0 = 50$. The rectangular yellow contours were added to guide the eye.

It is known that the roughening transition can only occur on certain crystallographic facets. For the studied BCC lattice, we did not find this effect on the (111) facet even in the warm regime ($\alpha = 1.3$ and $p = 0.69$). On the facets of (100)-type this process is observed, (see Fig. 6) although it is accompanied by a very small decrease in the average number of bonds, $\langle n_b(t) \rangle$, which is a qualitatively explainable physical effect in this case. The total surface area of the slab, $A_{slab}$, increases in time, and it would seem that $\langle n_b(t) \rangle$ should decrease. However, the slopes of the formed hillocks and cavities are made up of "scales" of the (110)-facets with a dense packing of atoms and a minimum of surface energy density, $\sigma$. As a result, the growth of $A_{slab}(t)$ is, up to some extent, compensated by a decrease in the average value of $\langle \sigma(t) \rangle_{A_{slab}}$. If the nanowire is oriented along the [100]-axis, then at the initial stage of its evolution it is bounded not only by the (110)-type planes (see Fig. 2), but also, strictly speaking, by (001)-facets. However, the contribution of these facets to the value of $A_{nw}$ is insignificant, since the distance between the nanowire axis and the (100)-planes exceeds this distance for the (110)-planes by $\sigma_{(001)}/\sigma_{(110)}$ times.

Let us consider the cases of orientations of the nanowire axis along other directions with low Miller indices - [110] and [111] - when the axis of the wire is the axis of its symmetry.

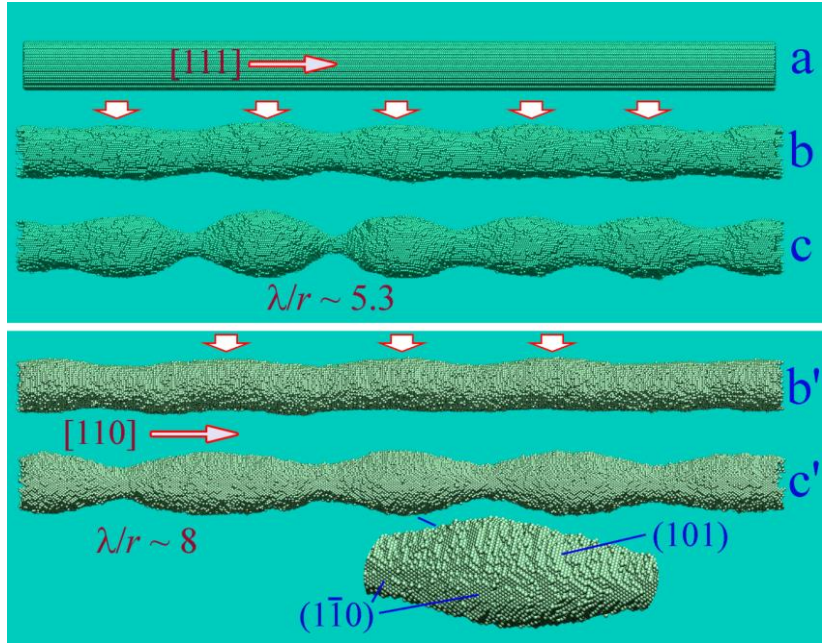

**Fig. 8. Dynamics of BCC nanowires with the [111]- and [110]-orientations in the hot regime.** $\alpha = 1.3$ and $p = 0.69$; $L = 630$, $d_0 = 40$, and $N_0 \approx 198 \times 10^3$ atoms. Evaporation is blocked. Sub-image (a) depicts the initial shape of the [111] BCC nanowire. Sub-images (b) and (c) represent the morphology of its surface at $t = 2.4$ and $5.25$ ($\times 10^6$) MC time steps, respectively. Sub-images (b′) and (c′) represent the shapes of the nanowire with [110]-orientation at $t = 4.5$ and $10$ ($\times 10^6$) MC time steps, respectively. The inset shows the morphology of the bulging region.

In the [110]-orientation, the lateral surface of the nanowire at the initial stage of evolution is bounded by only two planes of the (110)-type (see Fig. 7B). The rest of the area is formed by four facets of the (111)-type. It can be assumed in advance that the length of the perturbations, λ, during the development of the instability will lie in the range of $\lambda_{cr} < \lambda < \lambda_{max}$, i.e., $2\pi < \lambda/r_{nw} < 9$. On the one hand, the contribution of the (110)-facets to the total surface area is of the order of 1/3, so that the roughening transition effect is not dominant in the determination of λ; hence, the short-wave perturbations, $\lambda < \lambda_{cr}$, are hardly expectable. On the other hand, the slopes of the formed neck regions approach four facets of the (110)-type (see Fig. 7A), which intersect the axis of the wire at an angle $\theta = 30^o$. Correspondingly, the density of surface energy in newly formed sections of the surface will decrease with time (according to [75], the surface energy decreases linearly with decreasing angle between the (hkl)-planes and the (110)-plane). This factor stimulates shorter-wave perturbations ($\lambda < \lambda_{max}$) when compared with the case of an isotropic σ [31].

The case of (111)-orientation is quite simple for predicting the result. The surface of such a wire is transformed to a surface bounded by six faces of the (110)-type on which the roughening transition develops. Therefore, the expected breakup parameter, $\lambda/r_{nw}$, should be close to but slightly greater than in the case of [100]-orientation of the nanowire axis ($\lambda/r_{nw} \sim 4.5 - 5$, see Fig. 2). Bases for the prediction are the next conceptions. The roughening transition processes arising at the (110)-facets, which bound a nanowire at the initial stage, must be self-consisted in forming the slopes of necking regions to satisfy the required thermodynamical relations. In the case of (100)-orientation of a nanowire, four faces of the (110)-type with a minimum surface energy density cross the axis of the nanowire at an angle of $45^0$ and determine the formation of these necking regions (see Fig, 7A). In the (111)-orientation, only three (110) faces cut the wire at an angle of $\sim 45^0$. This difference in morphology determines an increase in $\lambda$ to minimize the growth of the total surface of the nanowire, $A_{nw}(\lambda; t)$, with the development of short-wave perturbations ($\lambda < \lambda_{cr}$). Data presented in Fig. 8 confirm the above results for the qualitative analysis of the discussed processes: $\lambda/r_{nw} \sim 5.3$.

### 3.2 Stimulated roughening transition

This section attempts to explain the interesting results obtained in experiments [32] with ultra-thin gold nanowires ($d_0 < 10\ nm$) − face centered crystal lattice (FCC). Note that temperatures $T > 200°C$ are sufficient for the development of Rayleigh instability of a nanowire with a diameter of 25 nm [38]. The dynamics of gold nanowires has been studied in many works. Their results show that, depending on the orientation of the nanowire relative to its internal crystal structure, the wavelength of developing surface modulations is often close to the classical value, $\lambda \sim \lambda_{max} = 9r_{nw}$, but significant deviations from the predictions of the theory up to $\lambda \sim (25 - 30)r_{nw}$ are also often observed. The result presented in [32] is surprising in the sense that in a nanowire irradiated with an electron beam, the wavelength of the excited modulations of the nanowire radius decreases to at least $\lambda \sim 5.5 r_{nw} < \lambda_{cr}$ (according to our estimates of the data presented in [32]) depending on the density electron beam energy. The amplitude of the modulations reaches a maximum in a few minutes (~ 10 min) and then the nanowire configuration becomes frozen in time. The minimum radius of curvature of the surface in the regions of its narrowing and the time of reaching to the quasi-stationary state sharply decrease with increasing beam intensity. The authors note moderate heating of the wire when irradiated with electrons (~$100°C$) and point to the "size effect" - owing to its small size, the nanowire modifications appear due to weak heating. The observed features of the shape transformation, "which is different from that observed with metallic nanowires with a diameter larger than 10 nm [32]", are associated with the intensification of surface diffusion of atoms under the action of irradiation.

Our interpretation, presented below, correlates with the conceptions presented by authors of [32] that consider an increase in the surface diffusion coefficient as a main factor in nanowire modifications. The interaction of surface atoms with beam electrons leads to a decrease in the activation energy of jumps, $\Delta$, and accordingly, to an increase in their frequency (parameter $p$ in our model). However, as it is shown in sequel, the observed effects in [32] can be realized only with a certain orientation of the nanowire axis, namely the [110]-direction. In this case, the lateral surface of the nanowire (at the initial stage of evolution) is mainly bounded by four (111)-type planes, which correspond to the minimum surface energy density in the FCC crystal lattice. On the two additional $(00 \pm 1)$-faces, $\sigma$ is also relatively low [48,74]. For this reason, nanowires with this orientation are the most resistant to breakup [34, 38, 44]. Naturally, on the slopes of developing necking regions, the value of $\sigma$ is higher than on the lateral surface of the bulges. Therefore, the maximum increment of the instability development shifts toward the wavelengths $\lambda > \lambda_{max}$, which correspond to a larger reduction in the total surface area of the nanowire, $A_{nw}$. In our work [48], in the so-called "cold regime" ($\alpha = 0.96$ and $p = 0.71$ for the FCC lattice), noticeable modulations of the nanowire radius with a wavelength of $\lambda/r_{nw} \sim 14.5$ ($r_{nw} = 10$) arose at $t \approx 8 \times 10^6$ MC steps (In the case under consideration, the unit of length, as before, is $a/2$, where $a$ is the FCC lattice constant). Accordingly, with a radius $r_{nw} = 17$, the decay time would increase significantly (see Fig. 9).

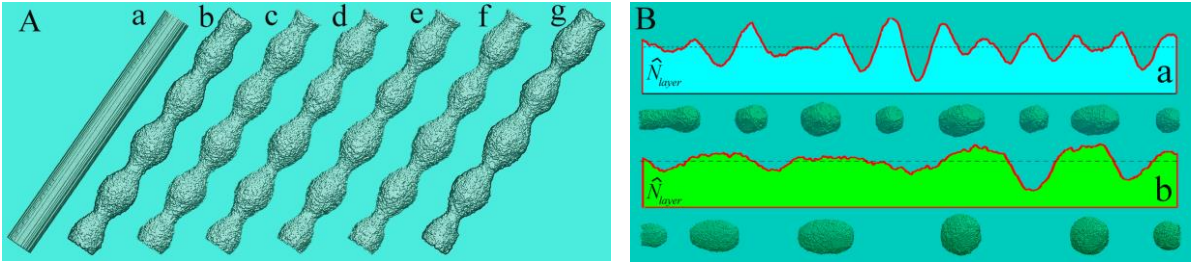

**Fig. 9**. **Diversity of surface transformations of nanowires with the FCC lattice structure and different orientations when the surface diffusion of atoms is accelerated by external treatments.** **(A)** [110]-orientation, low temperature regime with turned on evaporation: $\alpha = 1, p = 0.85, L = 350, d_0 = 34$ and $\lambda/r_0 \approx 4.6$. The nanowire contains $N_t \approx 160 \times 10^3$ atoms. Sub-images (a), (b), (c), (d), (e), (f), and (g) show the stabilized in time short-wave nanowire configurations after $t = 0.0, 0.36, 0.72, 1.08, 1.44, 1.8$ and $2.16$ ($\times 10^6$) MC time steps, respectively. **(B)** [111]-orientation, warm regime: $\alpha = 0.9$ and $p = 0.725$; $L = 720, d_0 = 24$, $N_t \approx 159 \times 10^3$; sub-image (a) depicts the distribution of the number of atoms in the atomic layers of (111)-type along the nanowire, $\widehat{N}_{layer} = N_{layer}/\langle N_{layer} \rangle$, at $t = 2.25 \times 10^6$ and the configuration of the nanowire at the final stage of its breakup, $t = 4.5 \times 10^6$ MC steps; sub-image (b) shows the long-wave nanowire disintegration for larger values of parameter $p$ ($p = 0.8, d_0 = 25$, and $N_t \approx 169 \times 10^3$) at times $t = 3.9 \times 10^6$ and $t = 7.8 \times 10^6$ MC steps, respectively.

Figure 9 presents the results of our simulation of the nanowire dynamics in the case of a moderate temperature ($\alpha = 1$), taking into account that heating by an electron beam is low. The intensification of surface diffusion is reflected in an increase in the parameter $p$ ($p = 0.85$; for $\alpha = 1$ in the model [45, 46, 67–73], $p = 0.7$). The diameter of the nanowire, $d_0 = 34$, in the case of gold corresponds to 6.9 nm. It can be seen that the saturation of the amplitude of the short-wavelength modulations of the nanowire radius (see sub-image (b) in Fig. 9A; $\lambda/r_{nw} \approx 4.6$) is achieved in a substantially short period of time, $t_s = 0.36 \times 10^6$, compared with the time of its breakup in the absence of irradiation with electrons.

The configuration of the nanowire is practically unchanged up to the time moment $t \approx 6t_s$. Freezing the evolution of a nanowire upon short-wavelength excitations, as we have already noted above, is a direct analogue of the equilibrium modulated by the radius configurations of infinite liquid filaments - unduloids [78]. In our case, such configurations eventually collapse as a result of the merger of single beads **[47]**.

The reduction of the wavelength, $\lambda$, and the saturation time of the perturbation amplitude, $t_s$, can be managed by the value of the parameter $p$ for a fixed $\alpha = 1$. Thus, the parameter $\lambda/r_{nw} \sim 5.5$ observed in [32] is achieved when $p$ increases from 0.7 [45,46,48] to 0.77. Note that at $p = 0.7$, the perturbation wavelength is $\lambda \sim 24 r_{nw}$ [48].

Results shown in Fig. 9A were obtained taking into account the evaporation of atoms from the surface of the nanowire, which reflects the presence of a "knock on" effect in [32]. However, blocking this process (calculating the dynamics of a nanowire without evaporation) practically did not change the modifications of the wire (only a slight, ~ 7%, decrease in $\lambda$ is observed at the initial stage of perturbations arising, but this distinction gradually vanishes with time).

Here we note an important detail. The observed effects are not directly related to the heating of the wire by an electron beam. The main role is played by the intensification of surface diffusion. Without this intensification, significant heating of the wire in the "hot mode" ($\alpha = 0.8$ and $p = 0.752$) leads to the excitation of modulations of radius with a wavelength of $\lambda \sim 10 r_{nw}$ [48].

The role of the electron beam in the development of instability in a nanowire is completely ambiguous (see Fig. 9B). In the orientation of its axis along the [111]-direction, the side surface of the wire is bounded by six planes of the (110)-type with the highest energy density, $\sigma_{(110)}$, in the case of the FCC lattice. When the instability arises, the necking regions are formed by three faces of the (111)-type and three faces of the (100)-type with lower surface energy densities, $\sigma_{(110)} > \sigma_{(100)} > \sigma_{(111)}$ [74]. With such an anisotropy of $\sigma$, the wavelength of perturbations is somewhat lower than $\lambda_{max} \approx 9 r_{nw}$ (see configuration (a) in Fig. 9B). Decreasing the activation barrier ($p = 0.8$ instead of 0.725) sharply increases both the breakup time and the period of modulation of the nanowire cross section, $\lambda \sim 12 r_{nw} > \lambda_{max}$. That is, the result of the influence of surface diffusion intensification on the perturbation wavelength is completely opposite to the result shown in Fig. 9A, and dramatically depends on the orientation of the nanowire axis. Note that the effect of a significant increase of the wavelength of developing disturbances ($\lambda \gtrsim 12 r_{nw} > \lambda_{max}$) is indeed observed in recent experiments [79], in which the abnormally high diffusion of surface atoms is caused by bombardment of the Ag-nanowire by $Ar^+$ ions formed in low-temperature RF (radio frequency) plasma (the temperature of the nanowire did not exceed $100^o C$).

Systematizing the results, we note the following. In the absence of electron beam irradiation, the breakup of nanowires with an FCC lattice corresponds to the classical scenario of Rayleigh instability, when the development of the instability is accompanied by a decrease in its total surface energy, $E_s$. Naturally, the anisotropy of $\sigma$ makes corrections to the value of $\lambda$ depending on the value of $\sigma$ in the regions of narrowing and broadening, without fundamentally changing the dynamics of energy $E_s$. With a decrease in the activation threshold, $\Delta$, the wavelength, $\lambda$, noticeably increases in the [111]-orientation (see Fig. 9B, (b)); however, this elongation does not change the instability development scenario: the surface energy decreases as a result of more expressive reduction in $A_{nw}(\lambda)$ ($\Delta A_{nw}(\lambda) < 0$), which is accompanied by smaller $\sigma$ in the areas of narrowing than in the areas of broadening. With the [110]-orientation, the situation dramatically changes: the surface energy increases both due to an increase in $A_{nw}(\lambda)$ at $\lambda < \lambda_{cr}$ and due to an increase in the density of surface energy in the narrowing regions. In this case, at a fixed temperature of the nanowire, the energy for modifying its morphology is provided by the environment, which "pushes" the system to overcome the potential barrier that prevented the realization of "energetic instability", i.e. the possibility of breakup into single nanoclusters with a total surface energy that is finally less than the initial energy of the nanowire. In our study, the abovementioned modification of the surface morphology of a quasi-one-dimensional nanosystem before breakup is associated with the development of roughening transition, which is known for two-dimensional systems [56–66].

## 4. Conclusions

In this work, we demonstrated the physical mechanisms on the basis of which non-characteristic scenarios of the surface dynamics of quasi-one-dimensional systems observed in experiments can be realized. The aforementioned mechanisms are associated with the emergence of roughening transition on the faces bounding a nanowire/slab at the initial stage of its evolution. The dominance of roughening transition in the generation of periodic modulations of a nanowire is possible when in the formation of its lateral surface the appropriate faces prevail, which depends on the nanowire orientation.

As an example, such a predominance is realized for nanowires with a body-centered crystal lattice in the cases when they are oriented in the [100]- and [111]-directions. In such orientations, their lateral surfaces are mainly represented by [110]-facets with a minimum surface energy density, which provides the basis for the next, seemingly flawless, qualitative estimation of the wavelength, $\lambda$, of developing perturbations. Since the surface energy density on the slopes of forming constrictions is higher than on the surface of broadening regions, such $\sigma$-anisotropy (called "positive" here) should be compensated by a more noticeable decrease in the area of the nanowire lateral surface than in the case of an isotropic value of $\sigma$. That is, the supposed decrease in surface energy, $E_s$, with the development of instability requires the fulfillment of the condition $\lambda > \lambda_{max} \approx 9r_{nw}$. However, the dynamics of the nanowire "follows the trajectory" with a decrease in the free energy, $F$, of the system, and in the cases under consideration is accompanied by an increase in $E_s$ (a decrease in the average number of bonds $\langle n_b \rangle$). If the determining factor in the dynamics of a nanowire is roughening transition, then the wavelength of perturbations and the radius of nanowire are not directly related. Therefore, the breakup parameter, $\lambda/r_{nw}$, depends on the radius and can decrease even below the energetic instability threshold, $\lambda/r_{nw} \lesssim 4.5$, which leads to the formation of metastable states.

However, the initial formation of the lateral surface of a nanowire from facets of the same type does not guarantee its non-characteristic evolution. A nanowire with an FCC crystal lattice, which is oriented along the [110]-direction, is mainly bounded by (111)-facets. Such facets are most resistant to roughening transition at temperatures below the melting point. With the development of instability, the anisotropy of $\sigma$ corresponds to the positive type and, in the absence of roughening transition, the nanowire evolution is in agreement with the above qualitative estimate of the parameter $\lambda/r_{nw} > 9$: at low temperatures, this parameter reaches values of the order of 25 [**34,38,44,48**]**.** However, roughening transition can be stimulated by the bombardment of nanowire surface by electrons [**32**] or cold plasma ions. As a result of such a stimulation of the surface diffusion of atoms, metastable states again arise with the parameter $\lambda/r_{nw} < 2\pi$.

Here we take note of an important fact: in experimental studies, a weak heating of the nanowire during its bombardment is noted. In our numerical simulations, the results of which are in good agreement with the experimental data [**32,79**], the activation of surface diffusion is presented by some increase in the parameter $p$, while the value of $\alpha$ corresponds to the weak heating of the nanowire. It is important to note that the acceleration of surface diffusion is not followed by nanowire heating. Otherwise, if for increased $p$ we use the value of $\alpha$ calculated according to Eq. (5) (as if heating is turned on), then the effects of external bombardment are sharply weakened. Thus, a variety of periodically-modulated quasi-one-dimensional configurations can be obtained by varying the orientation of the nanowire and methods of stimulating the surface diffusion of atoms without allowing significant heating of the nanowire.

**Conflicts of interest**
There are no conflicts to declare.

**Acknowledgements**
This work was supported by the Ministry of Education and Science of Ukraine (Project F2211).

# Supplementary information

# Roughening transition as a driving factor in the formation of self-ordered one-dimensional nanostructures

Vyacheslav N. Gorshkov[1], Vladimir V. Tereshchuk[1], and Pooya Sareh[2*]

[1] National Technical University of Ukraine, Igor Sikorsky Kyiv Polytechnic Institute,
37 Prospect Peremogy, Kiev 03056, Ukraine.

[2] Creative Design Engineering Lab (Cdel), School of Engineering, University of Liverpool, Liverpool, The Quadrangle, Brownlow Hill, L69 3GH, United Kingdom. *Corresponding author. Email: pooya.sareh@liverpool.ac.uk

## Supplementary text

As a result of surface diffusion of atoms at elevated temperatures, the roughly-cylindrical wire at the initial stage of its transformation takes on a shape bounded by planes with minimal surface energy densities. Fig. S1 demonstrates such a configuration for a nanowire with a radius $r_{nw} = 30$ achieved at moderate temperature: $\alpha = 1.5$ and $p = 0.65$. The lateral surface of the nanowire at a given time is mainly formed by bands of (110)-type faces, onto which there is a stream of atoms from intermediate zones (narrower bands of (100)-type faces).

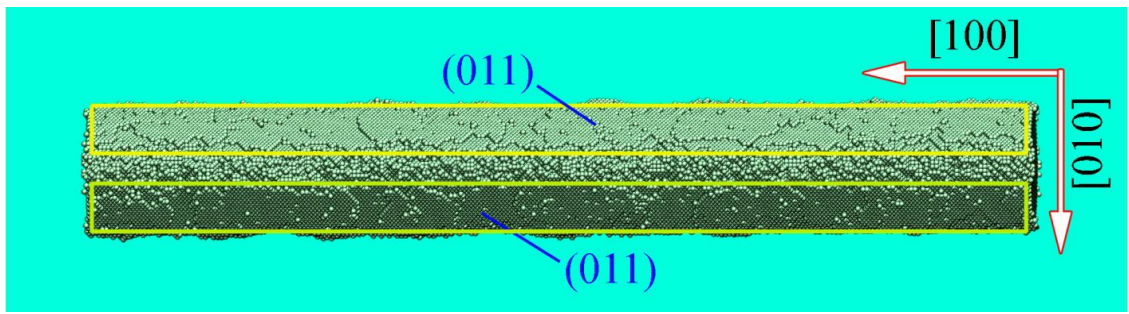

**Fig. S1.** Configuration of a nanowire with a body-centered lattice oriented along the [100]-direction at the initial stage of its evolution: $\alpha = 1.5$ and $p = 0.65$; $r_{nw} = 30$.

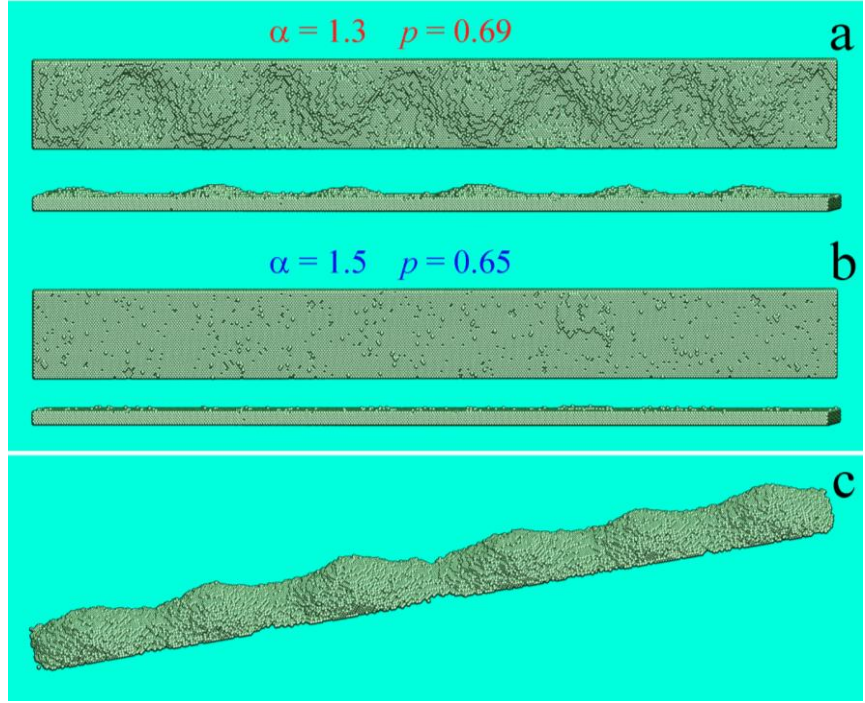

**Fig. S2**. Morphology of a slab with the upper (110) face after time $t = 10^7$ MC steps. $L = 630$, $w = 70$, and $h = 10$. Sub-images (a) and (b) show a modification of the slab that is "immersed" in a substrate of "frozen"/moveless atoms. Configuration (c) shows the shape of the plate, which is placed on the substrate: $\alpha = 1.5$ and $p = 0.65$.

This "external" stream of atoms can stimulate the formation of stepped hillocks on (110) faces. In the absence of external atomic flows, a pronounced modification of the (110)-type face (roughening transition) is possible only at a temperature exceeding the threshold. The threshold character of development of the roughening transition is shown in Fig. S2. Sub-images (a) and (b) represent the modification of the surface of a thin slab after time $t = 10^7$ MC steps. In this case, it is assumed that the slab is a part of a substrate with a (110) outer face. However, only the atoms that make up the plate are movable. The boundary condition for the movable atoms is reflective boundaries installed along the outer perimeter of the slab.

Data presented in Fig. S2 indicate that there is some intermediate value $1.5 > \alpha_R > 1.3$, which corresponds to the critical temperature, $T_R$, necessary for the occurrence of roughening transition. If the slab is not immersed in the substrate, but lies on it, then with a sufficiently small thickness it can break up into single fragments with a stepped structure. With a thicker slab, periodic perturbations of its surface occur in a two-mode regime.

At the initial stage of evolution, intense atomic fluxes from the side edges to the central part excite short-wavelength modulations of the slab height. As the edges are rounded and fast streams relax, longer wavelength modulations begin to play a dominant role (see video file Slab_on-Substrate.avi).

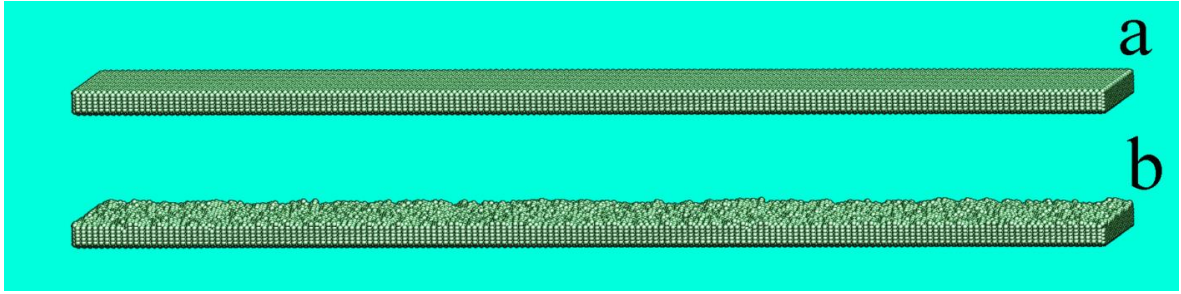

**Fig. S3**. Modification of the surface of the slab, the upper face of which is a (111)-facet; $\alpha = 1.3$ and $p = 0.69$. The geometrical dimensions of the plate are the same as in Fig. S1.

The roughening transition effect occurs only on faces of a certain type. Fig. S3 shows that only short-scale perturbations of a small-amplitude surface are observed on the (111) face — strictly ordered long-wavelength surface modulations do not develop.

# References

1. Roy, A.; Pandey, T.; Ravishankar, N.; Singh, A. K. Single crystalline ultrathin gold nanowires: Promising nanoscale interconnects. *AIP Adv*. **2013**, *3*, 032131
2. Hong K, Xie M, Hu R, Wu H: Diameter control of tungsten oxide nanowires as grown by thermal evaporation. *Nanotechnology* 2008, 19: 8.
3. Gu Z, Li H, Yang W, Xia Y, Yao J: Large-scale synthesis of single-crystal hexagonal tungsten trioxide nanowires and electrochemical lithium intercalation into the nanocrystals. *J Solid State Chemistry* 2007, 180: 1. 10.1016/j.jssc.2006.11.021
4. Pei, T., Thiyagu, S. and Pei, Z., 2011. Ultra high-density silicon nanowires for extremely low reflection in visible regime. *Applied Physics Letters*, 99(15), p.153108.
5. A. Najar, J. Charrier, P. Pirasteh, and R. Sougrat, "Ultra-low reflection porous silicon nanowires for solar cell applications," Opt. Express 20, 16861-16870 (2012)
6. Fonoberov, V. and Balandin, A. (2006). Giant Enhancement of the Carrier Mobility in Silicon Nanowires with Diamond Coating. *Nano Letters*, 6(11), pp.2442-2446.
7. Cezar B. Zota, David Lindgren, Lars-Erik Wernersson, and Erik Lind. Quantized Conduction and High Mobility in Selectively Grown $In_xGa_{1-x}As$ Nanowires. *ACS Nano* 2015, *9* (10), 9892-9897.
8. Hyungjun Lee and Hyoung Joon Choi. Single-Impurity Scattering and Carrier Mobility in Doped Ge/Si Core−Shell Nanowires. *Nano Letters* 2010, *10* (6), 2207-2210. https://doi.org/10.1021/nl101109p
9. Niquet, Y. and Delerue, C. (2012). Carrier mobility in strained Ge nanowires. *Journal of Applied Physics*, 112(8), p.084301.
10. Li, Yingfeng & Fu, Pengfei & Li, Ruike & Song, Dandan & Shen, Chao & Zhao, Yan. (2015). A comparison of light-harvesting performance of silicon nanocones and nanowires for radial-junction solar cells. Scientific reports. 5. 11532. 10.1038/srep11532.

11. Song, Tao & Lee, Shuit-Tong & Sun, Baoquan. (2012). Silicon nanowires for photovoltaic applications: The progress and challenge. Nano Energy. 1. 654–673. 10.1016/j.nanoen.2012.07.023.
12. Wallentin J., Anttu N., Asoli D., Huffman M., Åberg I., Magnusson M. H., Siefer G., Fuss-Kailuweit P., Dimroth F., Witzigmann B., Xu H. Q., Samuelson L., Deppert K., Borgström M. T. (2013). InP nanowire array solar cells achieving 13.8% efficiency by exceeding the ray optics limit. *Science 339, pp.* 1057–1060.
13. Ruffo, R., Hong, S., Chan, C., Huggins, R. and Cui, Y. (2009). Impedance Analysis of Silicon Nanowire Lithium Ion Battery Anodes. *The Journal of Physical Chemistry C*, 113(26), pp.11390-11398.
14. Wang, W., Chen, C., Lin, K., Fang, Y. and Lieber, C. (2005). Label-free detection of small-molecule-protein interactions by using nanowire nanosensors. *Proceedings of the National Academy of Sciences*, 102(9), pp.3208-3212.
15. Stern, E., Vacic, A., Rajan, N., Criscione, J., Park, J., Ilic, B., Mooney, D., Reed, M. and Fahmy, T. (2009). Label-free biomarker detection from whole blood. Nature Nanotechnology, 5(2), pp.138-142.
16. Chua, J., Chee, R., Agarwal, A., Wong, S. and Zhang, G. (2009). Label-Free Electrical Detection of Cardiac Biomarker with Complementary Metal-Oxide Semiconductor-Compatible Silicon Nanowire Sensor Arrays. Analytical Chemistry, 81(15), pp.6266-6271.
17. C. Jagadish, "Semiconductor Nanowires for Optoelectronics Applications," in *Asia Communications and Photonics Conference (ACPC) 2019*, OSA Technical Digest (Optical Society of America, 2019), paper S2A.1.
18. Day, R., Mankin, M., Gao, R., No, Y., Kim, S., Bell, D., Park, H. and Lieber, C., 2015. Plateau–Rayleigh crystal growth of periodic shells on one-dimensional substrates. *Nature Nanotechnology*, 10(4), pp.345-352.
19. Day, R., Mankin, M. and Lieber, C., 2016. Plateau–Rayleigh Crystal Growth of Nanowire Heterostructures: Strain-Modified Surface Chemistry and Morphological Control in One, Two, and Three Dimensions. *Nano Letters*, 16(4), pp.2830-2836.
20. You, G., Gong, H. and Thong, J., 2010. Improving the morphological stability of a polycrystalline tungsten nanowire with a carbon shell. *Nanotechnology*, 21(19), p.195701.
21. Zhang, Y., Yan, Y. and Zhu, F., 2007. The Periodic Instability of Diameter of ZnO Nanowires via a Self-oscillatory Mechanism. *Nanoscale Research Letters*, 2(10), pp.492-495.
22. Xue, Z., Xu, M., Zhao, Y., Wang, J., Jiang, X., Yu, L., Wang, J., Xu, J., Shi, Y., Chen, K. and Roca i Cabarrocas, P. (2016). Engineering island-chain silicon nanowires via a droplet mediated Plateau-Rayleigh transformation. *Nature Communications*, 7(1).
23. Agati, M., Boninelli, S., Castrucci, P., Amiard, G., Pandiyan, R., Kolhatkar, G., Dolbec, R., Ruediger, A. and El Khakani, M. (2018). Formation of silicon nanocrystal chains induced via Rayleigh instability in ultrathin Si/SiO2 core/shell nanowires synthesized by an inductively coupled plasma torch process. *Journal of Physics: Materials*, 2(1), p.015001.
24. Chu, Y., Jing, S., Yu, X. and Zhao, Y. (2018). High-Temperature Plateau–Rayleigh Growth of Beaded SiC/SiO2 Nanochain Heterostructures. *Crystal Growth & Design*, 18(5), pp.2941-2947.
25. Takasaki, M., Tago, M., Oaki, Y. and Imai, H., 2020. Thermally induced fragmentation of nanoscale calcite. *RSC Advances*, 10(10), pp.6088-6091.
26. Wang, H., Upmanyu, M. and Ciobanu, C. (2008). Morphology of Epitaxial Core−Shell Nanowires. Nano Letters, 8(12), pp.4305-4311.

27. Xue, Z., Xu, M., Li, X., Wang, J., Jiang, X., Wei, X., Yu, L., Chen, Q., Wang, J., Xu, J., Shi, Y., Chen, K. and Roca i Cabarrocas, P. (2016). In-Plane Self-Turning and Twin Dynamics Renders Large Stretchability to Mono-Like Zigzag Silicon Nanowire Springs. *Advanced Functional Materials*, 26(29), pp.5352-5359.
28. Gloppe, A. *et al.* Bidimensional nano-optomechanics and topological backaction in a non-conservative radiation force field. *Nature Nanotech.* **9,** 920–926 (2014).
29. Ramos, D. *et al.* Optomechanics with silicon nanowires by harnessing confined electromagnetic modes. *Nano Lett.* **12,** 932–937 (2012).
30. J. Plateau, *Experimental and Theoretical Statics of Liquids Subject to Molecular Forces Only*, **1** (Gauthier-Villars, Paris, 1873).
31. Nichols F. A., Mullins W. W. (1965). Surface- (interface-) and volume-diffusion contributions to morphological changes driven by capillarity. *Trans. Metal. Soc. AIME* 233, pp. 1840–1848.
32. Xu, S., Li, P. and Lu, Y., 2017. In situ atomic-scale analysis of Rayleigh instability in ultrathin gold nanowires. *Nano Research*, 11(2), pp.625-632.
33. Li, H., Biser, J., Perkins, J., Dutta, S., Vinci, R. and Chan, H. (2008). Thermal stability of Cu nanowires on a sapphire substrate. *Journal of Applied Physics*, 103(2), p.024315.
34. Rauber, M., Muench, F., Toimil-Molares, M. and Ensinger, W. (2012). Thermal stability of electrodeposited platinum nanowires and morphological transformations at elevated temperatures. *Nanotechnology*, 23(47), p.475710.
35. Volk, A., Knez, D., Thaler, P., Hauser, A., Grogger, W., Hofer, F. and Ernst, W. (2015). Thermal instabilities and Rayleigh breakup of ultrathin silver nanowires grown in helium nanodroplets. *Physical Chemistry Chemical Physics*, 17(38), pp.24570-24575.
36. Gill, S. (2013). Controlling the Rayleigh instability of nanowires. *Applied Physics Letters*, 102(14), p.143108.
37. Toimil Molares, M., Balogh, A., Cornelius, T., Neumann, R. and Trautmann, C. (2004). Fragmentation of nanowires driven by Rayleigh instability. *Applied Physics Letters*, 85(22), pp.5337-5339.
38. Karim, S., Toimil-Molares, M., Balogh, A., Ensinger, W., Cornelius, T., Khan, E. and Neumann, R. (2006). Morphological evolution of Au nanowires controlled by Rayleigh instability. *Nanotechnology*, 17(24), pp.5954-5959.
39. Karim, S., Toimil-Molares, M., Ensinger, W., Balogh, A., Cornelius, T., Khan, E. and Neumann, R. (2007). Influence of crystallinity on the Rayleigh instability of gold nanowires. *Journal of Physics D: Applied Physics*, 40(12), pp.3767-3770.
40. Gurski, K. and McFadden, G. (2003). The effect of anisotropic surface energy on the Rayleigh instability. *Proceedings of the Royal Society of London. Series A: Mathematical, Physical and Engineering Sciences*, 459(2038), pp.2575-2598.
41. Cahn, J. (1979). Stability of rods with anisotropic surface free energy. *Scripta Metallurgica*, 13(11), pp.1069-1071.
42. Min, D. and Wong, H. (2006). Rayleigh's instability of Lennard-Jones liquid nanothreads simulated by molecular dynamics. *Physics of Fluids*, 18(2), p.024103.
43. Zhao, C., Sprittles, J. and Lockerby, D. (2019). Revisiting the Rayleigh–Plateau instability for the nanoscale. *Journal of Fluid Mechanics*, 861.

*44.* Vigonski, S., Jansson, V., Vlassov, S., Polyakov, B., Baibuz, E., Oras, S., Aabloo, A., Djurabekova, F. and Zadin, V. (2017). Au nanowire junction breakup through surface atom diffusion. *Nanotechnology*, 29(1), p.015704.
45. Gorshkov, V., Sareh, P., Tereshchuk, V. and Soleiman- Fallah, A. (2019). Dynamics of Anisotropic Break- Up in Nanowires of FCC Lattice Structure. *Advanced Theory and Simulations*, pp.1900118.
46. Gorshkov, V. and Privman, V. (2017). Kinetic Monte Carlo model of breakup of nanowires into chains of nanoparticles. *Journal of Applied Physics*, 122(20), p.204301.
47. Gorshkov, V., Tereshchuk, V. and Sareh, P. (2019). Restructuring and breakup of nanowires with the diamond cubic crystal structure into nanoparticles. *Materials Today Communications*, p.100727.
48. Gorshkov, V., Tereshchuk, V. and Sareh, P., 2020. Diversity of anisotropy effects in the breakup of metallic FCC nanowires into ordered nanodroplet chains. *CrystEngComm*, 22(15), pp.2601-2611.
49. Li, P., Han, Y., Zhou, X., Fan, Z., Xu, S., Cao, K., Meng, F., Gao, L., Song, J., Zhang, H. and Lu, Y., 2020. Thermal Effect and Rayleigh Instability of Ultrathin 4H Hexagonal Gold Nanoribbons. *Matter*, 2(3), pp.658-665.
50. Zugarramurdi, A., Borisov, A., Zabala, N., Chulkov, E. and Puska, M., 2011. Clustering and conductance in breakage of sodium nanowires. *Physical Review B*, 83(3).
51. Lai, K. and Evans, J., 2019. Reshaping and sintering of 3D fcc metal nanoclusters: Stochastic atomistic modeling with realistic surface diffusion kinetics. *Physical Review Materials*, 3(2).
52. Wong, H., 2012. Energetic instability of polygonal micro- and nanowires. *Journal of Applied Physics*, 111(10), p.103509.
53. Kim, G. and Thompson, C., 2015. Effect of surface energy anisotropy on Rayleigh-like solid-state dewetting and nanowire stability. *Acta Materialia*, 84, pp.190-201.
54. Kubendran Amos, P., Mushongera, L., Mittnacht, T. and Nestler, B., 2018. Phase-field analysis of volume-diffusion controlled shape-instabilities in metallic systems-II: Finite 3-dimensional rods. *Computational Materials Science*, 144, pp.374-385.
55. Vu Thien Binh, Uzan, R. and Drechsler, M., 1978. The surface diffusion of tungsten at very high temperatures. Journal de Physique Lettres, 39(21), pp.385-388.
56. Claudin, P., Durán, O. and Andreotti, B., 2017. Dissolution instability and roughening transition. *Journal of Fluid Mechanics*, 832.
57. Physics, C., Desjonqueres, M., Spanjaard, D. and Heidelberg, S., 2020. *Concepts In Surface Physics | M-C. Desjonqueres | Springer*. [online] Springer.com. Available at: <https://www.springer.com/gp/book/9783642974847> [Accessed 20 May 2020].
58. Weeks J.D. (1980) The Roughening Transition. In: Riste T. (eds) Ordering in Strongly Fluctuating Condensed Matter Systems. NATO Advanced Study Institutes Series (Series B: Physics), vol 50. Springer, Boston, MA.
59. Maxson, J., Savage, D., Liu, F., Tromp, R., Reuter, M. and Lagally, M. (2000). Thermal Roughening of a Thin Film: A New Type of Roughening Transition. Physical Review Letters, 85(10), pp.2152-2155.
60. Andersen, M. and Ghoniem, N., 2007. Surface Roughening Mechanisms for Tungsten Exposed to Laser, Ion, and X-Ray Pulses. *Fusion Science and Technology*, 52(3), pp.579-583.
61. Scott, J., Hayward, S. and Miyake, M., 2005. High temperature phase transitions in barium sodium niobate: the wall roughening 1q–2q incommensurate transition and mean field tricritical behaviour in a disordered exclusion model. *Journal of Physics: Condensed Matter*, 17(37), pp.5911-5926.
62. Yu, J., Baldwin, M. and Doerner, R., 2017. Cracking and surface roughening of beryllium–tungsten alloy due to transient heating. *Physica Scripta*, T170, p.014009.
63. Heyraud, J., Métois, J. and Bermond, J. (1999). The roughening transition of the Si{113} and Si{110} surfaces – an in situ, real time observation. *Surface Science*, 425(1), pp.48-56.
64. Suzuki, T., Minoda, H., Tanishiro, Y. and Yagi, K. (1999). REM studies of the roughening transitions of Si high index surfaces. *Thin Solid Films*, 343-344, pp.423-426.
65. Dashti-Naserabadi, H., Saberi, A. and Rouhani, S. (2017). Roughening transition and universality of single step growth models in (2+1)-dimensions. *New Journal of Physics*, 19(6), p.063035.
66. Gao, H. and Nix, W., 1999. SURFACE ROUGHENING OF HETEROEPITAXIAL THIN FILMS. *Annual Review of Materials Science*, 29(1), pp.173-209.
67. V. Gorshkov, A. Zavalov, and V. Privman, Shape Selection in Diffusive Growth of Colloids and Nanoparticles, Langmuir 25, 7940–7953 (2009).
68. V. Gorshkov and V. Privman, Models of Synthesis of Uniform Colloids and Nanocrystals, Physica E 43, 1–12 (2010).
69. V. Privman, V. Gorshkov, and Y. E. Yaish, Kinetics Modeling of Nanoparticle Growth on and Evaporation off Nanotubes, J. Appl. Phys. 121, 014301-1–014301-8 (2017).
70. V. Gorshkov, V. Kuzmenko, and V. Privman, Nonequilibrium Kinetic Modeling of Sintering of a Layer of Dispersed Nanocrystals, CrystEngComm 16, 10395–10409 (2014).
71. V. Gorshkov, V. Kuzmenko, and V. Privman, Mechanisms of Interparticle Bridging in Sintering of Dispersed Nanoparticles, J. Coupled Syst. Multiscale Dynamics 2, 91–99 (2014).
72. V. Privman, V. Gorshkov, and O. Zavalov, Formation of Nanoclusters and Nanopillars in Nonequilibrium Surface Growth for Catalysis Applications: Growth by Diffusional Transport of Matter in Solution Synthesis, Heat Mass Transfer 50, 383–392 (2014).
73. V. Gorshkov, O. Zavalov, P. B. Atanassov, and V. Privman, Morphology of Nanoclusters and Nanopillars Formed in Nonequilibrium Surface Growth for Catalysis Applications, Langmuir 27, 8554–8561 (2010).
74. Wang, S., Tian, E. and Lung, C., 2000. Surface energy of arbitrary crystal plane of bcc and fcc metals. *Journal of Physics and Chemistry of Solids*, 61(8), pp.1295-1300.
75. Wen, Y. and Zhang, J., 2008. Surface energy calculation of the bcc metals by using the MAEAM. *Computational Materials Science*, 42(2), pp.281-285.
76. Galanakis, I., Bihlmayer, G., Bellini, V., Papanikolaou, N., Zeller, R., Blügel, S. and Dederichs, P., 2002. Broken-bond rule for the surface energies of noble metals. *Europhysics Letters (EPL)*, 58(5), pp.751-757.
77. Luo, Y. and Qin, R., 2014. Description of Surface Energy Anisotropy for BCC Metals. *Advanced Materials Research*, 922, pp.446-451.
78. Bertozzi, A. and Witelski, T. (1998). Axisymmetric Surface Diffusion: Dynamics and Stability of Self-Similar Pinchoff. *Journal of Statistical Physics*, 93(3/4), pp.725-776
79. Dai, H., Yu, X., Zhao, Z., Shi, D., Shi, X., Zhao, J., Dong, X. and Zhang, D., 2020. Low Temperature RF-Plasma Initiated Rapid and Highly Ordered Fracture on Ag Nanowires. *Applied Sciences*, 10(4), p.1338.